\documentclass[journal,twoside,12pt]{IEEEtran}
\usepackage[nocompress]{cite}
\usepackage{todonotes}
\usepackage{times}
\usepackage{color}
\usepackage{amsfonts,amssymb}
\usepackage{amsmath,bm}
\usepackage{ifpdf}
\usepackage{epsfig}
\usepackage[small]{subfigure}
\usepackage{fmtcount}
\usepackage[T1]{fontenc}
\usepackage{balance}
\usepackage{xcolor}
\usepackage{multirow}

\usepackage{perpage} %the perpage package
\MakePerPage{footnote} %the perpage package

\usepackage{array}
\usepackage{eqparbox}
\usepackage{color}
\usepackage{graphicx} 
\usepackage{url} 
\usepackage[small]{subfigure}
\usepackage{amsmath}
\usepackage{makecell}
\usepackage{balance}
\usepackage{tabu}
\usepackage{algorithm}
\usepackage{algorithmic}
\usepackage{longtable}

\usepackage{pifont}% http://ctan.org/pkg/pifont
\usepackage[utf8]{inputenc}
\usepackage[T1]{fontenc}
\usepackage{textcomp}
\usepackage{xcolor}

\usepackage{graphicx} % for rotatebox
\usepackage{wasysym} % for circles
\usepackage[utf8]{inputenc}
\newcommand*\rot{\rotatebox{90}}
\newcommand*\tableY{\CIRCLE}
\newcommand*\tableP{\RIGHTcircle}
\newcommand*\tableN{\Circle}

\def\BibTeX{{\rm B\kern-.05em{\sc i\kern-.025em b}\kern-.08em
        T\kern-.1667em\lower.7ex\hbox{E}\kern-.125emX}}

\begin{document}

\title{Reviewing Best Practices in Online Conferencing}

  \author{Simone Ferlin, Oliver Hohlfeld, Vaibhav Bajpai
  
  %\thanks{S. Ferlin is with Ericsson AB, Sweden}
  %\thanks{V. Bajpai is with CISPA Helmholtz Center for Information Security}
  %\thanks{O. Hohlfeld is with Brandenburg University of Technology}
  %\thanks{J. Crowcroft is with \dots}
  %\thanks{S. Keshav is with \dots}

  }

\maketitle

\begin{abstract}
  The COVID-19 pandemic disrupted the usual ways the networking research community operates. This article reviews experiences organising and participating in virtual conferences during the COVID-19 pandemic between 2020-2021. Thanks to the broader scope of the Dagstuhl seminar on 'Climate Friendly Internet Research' held in July 2021, here we focus the discussion on state-of-the-art in technologies and practices applied in online events such as conferences, teaching, and other meetings and identify approaches that are successful as well as others that need improvement. We also present a set of best practices and recommendations for the community.
\end{abstract}
%which set itself the broader goal of identifying approaches to reducing the carbon footprint from travel in the networking community



\section{Introduction}

The COVID-19 pandemic acknowledged once again how the Internet greatly facilitates remote collaboration. Yet, when it comes to research community activities, there is a certain culture of in-person only and mandatory participation. That means, conducting research indirectly results into a certain expectation around travelling and a high carbon footprint~\cite{fraser16}. One way to reduce this footprint is by means of how communities are structured and operated, e.g., many commitments part of keeping the community running could be online. 

During the COVID-19 pandemic, we witnessed a rapid transition to a virtual operation mode, which included remote working with online meetings and online community events. Then, the pandemic served as a catalyst to bring virtual events more into the mainstream. Due to the circumstances, many research communities experienced first-hand how it is to continue with its activities and carry out research without the need of travelling. Also, some impulse came in the direction of quantifying the impact in the carbon footprint of the way communities and activities are run: The author in~\cite{migault21} simulates the carbon footprint of the Internet Engineering Task Force (IETF) community, and ~\cite{OBRINGER2021105389} analyses the impact of video conferencing in the carbon footprint compared to metrics such as transportation.
%e.g., from publication and dissemination of research in conferences to peer-review in Technical Program Committee (TPC) meetings

To document the knowledge gained during the COVID-19 pandemic between 2020-2021, we evaluate experiences collected while running and participating in virtual events such as conferences, teaching and project meetings. We want to understand what was experienced by networking community members as more and less successful implementing and deploying virtual events, i.e., type of challenges encountered and which improvements are needed. Ultimately, this paper is part of the output of the Dagstuhl seminar on 'Climate Friendly Internet Research'\cite{bajpai_et_al:DagRep.11.6.14}, which covered a broader scope: The result of the seminar is a summary of recommendations to design hybrid events in~\cite{bajpai_et_al:CCR21}, where, this paper, focuses on reviewing best practices and lessons learnt in online events. %online events are part of the discussions carried out with all invited participants during the seminar.
To this end, we start looking at guidelines and best practices (2) followed by the various implications of online conferences (3). Then, we conclude with lessons learned (4) looking from both groups' perspective during the Dagstuhl seminar organising and participating in various types of online meetings, and our analysis of the discussions and takeaway (5).

\section{Guidelines and Best Practices}
\label{sec:guidelines:and:best:practices}

Here, we introduce guidelines for what we define as traditional aspects of conferences, e.g., agenda and time scheduling and research dissemination, and social aspects. To begin with, it is general consensus that online conferences are different from those organised as in-person events. As such, it should not be the goal that online conferences deliver the same experience as close to what an in-person conference would offer. Instead, in both formats, conference participants should be able to: 1) learn about and have access to research work, 2) meet other participants in some form of semi-structured interactions, and 3) be able to formally and informally present their own research. These three goals should be achieved using procedures and tooling available that may differ from traditional in-person conferences. Nevertheless, the outcome of such an investment for the online format should lead to the same outcome of an in-person event. In the following, we enlist the most critical aspects when rethinking in-person conferences as online events.

\subsection{Agenda and time scheduling} 
Online conferences require to revisit the traditional 20-30 minute paper presentation in a single timezone. Deciding about the schedule and distribution of the presentations to different timezones is perhaps the most critical aspect. During presentations, organisers should also monitor the scheduled agenda as well as the participants for any sort of technical issues, e.g., display of the slides, audio or video quality, etc. %The goal should be for the conference schedule to provide a suitable framework for all conference events, allowing participants to meet the "twin conference" goals of learning and interaction. 
There are three main aspects in the organisation, which we would like to highlight:
\vspace{-0.1pt}
\begin{itemize}
\item
  \emph{Different time zones:} Not all participants will be present at the same time in the online conference. Besides presentations, it is necessary to create mechanisms that allow interaction across time zones, e.g., how to maintain continuity for people who come in and out. Moreover, there may be only a limited number of hours where all participants can be present, where this time should be planned for plenary sessions such as keynote talks, best paper awards or some general announcements and interaction. We strongly recommend that the schedule shows participant-tailored timezones so that a participant finds easily where to be.
\item
  \emph{Zoom fatigue:} It is necessary to compress schedules into, perhaps, a four-hour limit for each day, and spread the event over more days. Also, we recommend that organizers reduce the number of papers presented per session, e.g., by selecting only the best ones determined already during the review process, instead of including every single paper. The remaining papers for the same session can be offered as posters or videos accessible at the participants' convenience. Another alternative is to create a multi-track conference, which has its own challenges, and it is still run online.
\item
  \emph{Ability to focus online:} From our experience, 8-minute talks with discussion sessions seem to have worked well, especially with pre-recorded materials. The talks are however limited to focus on the problem statement and main concluding aspects of the paper. Note that not everyone may appreciate this approach, as it may bring the participant to go through the entire paper, instead of getting an in depth understanding from the presentation.
\end{itemize}

\subsection{Navigation and signposting} 
The conference schedule should be easy to learn about the paper or topic handled in the session, join or rejoin events. This is even more important when sessions are now broken into smaller parts with fewer papers. Thus, signposting and a clearly structured landing page are needed. It would be helpful to directly jump to the session or breakout rooms with a single click. However, this requires deep link, which is currently not supported by all popular videoconferencing systems.
%i.e, users are sent directly to the application instead of landing in a registration or login website,
%Currently, some conferences' program websites are crowded with information, making it difficult to find information. 

\subsection{Poster sessions} 
In-person poster sessions allow participants to scan through several posters, with the option to start a conversation or just move to the next one. Here, online conferences have two main issues: 1) \emph{Social awkwardness}: It is not possible to quickly skim through the poster, making it rather awkward to just quickly join and leave. 2) \emph{Difficult navigation}: It is difficult to find posters within topics of interest or move across mixed mediums, e.g., from Zoom to Hubs - specialised spatial-metaphor tools. During this process, we gained experience with three solutions: \emph{Mozilla Hubs}~\cite{mozilla} allows for bi-modal feedback. \emph{Gather.town}~\cite{gather} makes it difficult to identify neighbors, i.e., author or another person standing next to the poster. Finally, \emph{Spatial.chat}~\cite{spatial} lacked good audio quality, and felt overall not fit for poster sessions. Some suggestions when using these tools are:

\begin{itemize}
\item \textit{Quick overview}: The ability of quick skim through the posters is important. There is need of improvement for the social aspects of it, e.g., openly state during the session that it is expected that participants enter and leave.
\item \textit{Breakout rooms}: One breakout room per poster during the poster session has shown to be the best configuration. This way, participants can "walk by" more freely, reducing the number of people per session, and allowing for a better interaction with the poster presenter.
\item \textit{Speed-dating}: Such approaches maximize the use of time. This has shown to be particularly helpful for younger research community members. Some conferences have also followed a so-called \textit{one minute madness} approach with the opportunity to arrange longer time slots with the presenter, if necessary.
\end{itemize}

\subsection{Structured chaos} 
Perhaps the most challenging aspect in online conferences is to replicate the social and hallway unstructured conversations. The main issue identified is how to get seed conversations going and conversations to further develop. There are several motivating factors, e.g., catch-up and strengthen existing bonds, making new introductions to senior participants, building the community and renewing existing relationships, identifying potential research partners, and, finally, recruitment and job hunting.
%It is also general consensus that these aspects have not been successfully addressed by current available tools.

Although we have not yet found a better tool for social interaction, there are some positive experiences: For short coffee breaks, dropping people into breakout rooms at random showed to work well. We think that there is need to explicitly identify \textit{social butterflies} who can actively promote social interaction and start off the conversations. This challenge mirrors quite a bit in experiences with online teaching, where breakout sessions for students face similar issues. Some difficulties and suggestions going through this process are:
\begin{itemize}
\item
  How to decide who should start conversations? A couple of examples of questions that can be used as conversations starters: ``Tell me about your work'' or ``You use a certain tool (or technology), what do you think of it'' or ``What have you been up to recently ?''.
\item 
  How to ensure implicit social behaviour is made explicit? There is a need to assign people certain roles so that they do not feel 
  awkward. For example, explicitly approach specific senior members of the community to steer the communication. At the same time, how to avoid toxic behavior and egotists? We need to make a careful selection of people for dedicated roles.
\item
  How to facilitate \textit{cross-pollination}? This needs to be made explicit, perhaps with a special newcomers meeting event with a chance to meet more senior members. To this end, how to strongly encourage senior members to join? Social interaction can be promoted by assigned seating in in-person meetings. However, the issue of how to balance people in meetings, i.e., half known or half newcomers, is still uncharted territory. How to (actively) bootstrap chaos? Perhaps this can be done using social enabler tools and senior members. 
\end{itemize}

\subsection{Text channel} 
A supporting text channel tool, usually Slack~\cite{slack}, emerged as a good idea for coordinating Question and Answers (Q\&A). Traditionally, Q\&A sessions provide additional context, or expose lacunae that reviewers did not catch. Such sessions can be intimidating for shy presenters, where a text channel-based Q\&A session allows participation, even anonymous. It also allows author responses to be better captured and shared with everyone, unlike the situation in a typical in-person conference. Some guidelines using Slack or similar text channel are:
\begin{itemize}
\item
  Session chairs need to be strict enforcing discipline in the Slack channel(s) to prevent discussions from wandering away. It is helpful to have a moderator or scribe assigned to capture the Q\&A content, also with the possibility to turn this content into a report to be published later (with consent). It might even be possible for scribes to report on "bits and bytes from the previous day" as in other communities. %such as Reseaux IP Europeens (RIPE) meetings.
\item
  The audience can be encouraged to make use of special signs to notify authors of pending questions, where authors should reply to them within 24 hours. A single Slack channel per session showed to work better, though, there is still a need to find questions for each paper. %On the other hand, one Slack channel per paper may get hard to navigate and difficult to find which channel to attend. 
\end{itemize}
An alternative to a text-channel is Slido~\cite{slido}, which allows questions to be posted and voted on, especially interesting for larger audiences and, thus, interesting for hybrid conferences as well.

\subsection{Audio, video and lighting issues}
For some participants, audio intelligibility is often better than in-person events, since it allows lip-reading, captioning, and per user defined audio settings. Nevertheless, bad audio and video quality continues to be a challenge: Audio issues are not only serious (`I have no audio'), but more subtle, such as issues with the microphone connection, background noise, echoing, as well as audio volume. Meanwhile, testing video submissions prior to the conference is a good idea, since there are still problems such as videos that do not work on certain operating system formats or require specialised players. Finally, lighting can be an issue, especially background-lighting, where light adjustment available in some systems such as GoogleMeet~\cite{meet}, help participants without prior guidance.
%An open research topic would be to use automation to judge quality of submitted videos. Alternatively, online conferences could just provide test sessions for interactive sessions such as panels and keynotes.

\section{Implications of Online Conferences}
\label{subsec:implications-online-conferences}

Now we present the financial, participation and diversity, as well as timezone aspects that mostly differ from in-person conferences.

\textbf{Costs and transparency --} There are several costs implied in online conferences, e.g., videoconferencing platforms such as Zoom, Google Meet, etc. Although many groups or universities already have licenses to these tools for teaching or faculty general usage other supporting tools are usually not part of this package. For example, Slack, social platforms (Gather.town or Spatial Chat), meeting give-aways (tokens and T-shirts), stenographers and real-life captions, simultaneous translations. %, and need to be considered separately.

Further, there are some hidden costs in the process of publishing, where established researchers begun to move to free research channels and platforms, e.g., \textit{arXiv}. Yet, younger researchers are expected to first publish in well-established venues, which imply paying fees. With the emergence of online conference models, financial and business models around research publishers will have to be reinvented~\cite{mosche21}: Some fraction of costs in in-person conference fees are related to its physical aspects, i.e., venue, technology support, and meals. Currently, there are several revenue streams, some of which are used for registration fees, sponsorships, and access to research papers. %Organisers could also help to promote environmental projects (which is good for reputation of the event.)

\textbf{Sponsorships} -- There seems to be more hesitancy in sponsoring online events, which used physical presence to attract talents as the motivation. Here, organisers need to consider facilitating conversations for recruiters, sponsors and peering coordinators with participants in online conferences.

\textbf{Travel funds --} It is unclear why and whether employers or funding agencies should fund travel and conference attendance when there is the possibility to attend online. Here, newcomers to the community have more trouble justifying the funding from now on. Thus, organisations need to rethink and re-purpose travel funds and scholarships.

\textbf{Conference local hubs --} Large community meetings, e.g., the IETF, are considering \textit{local hubs} in addition to the event that already runs online. The associated costs for such local hubs is presently unclear, however, such initiatives could also help including participants from low-income and less-represented groups to experience the community and build the local networking via the hub.

\textbf{Diversity --} Online conferences help improve diversity, encouraging participation from those who usually cannot afford travelling or have other issues, e.g., visa, family, health restrictions. Further, online archiving helps broader access to the conference materials. Meanwhile, smaller (local) events can also be run in a local spoken language, promote, and strengthen local communities. Some venues are also offering childcare for attendees, thus, also facilitating inclusion of another group of participants.
%Some large communities, e.g., RIPE, offer in addition stenography to help with inclusion. 

\textbf{Time zones --} The ACM SIGCOMM 2020 and 2021 followed both a model of pre-recorded presentations together with multiple Q\&A sessions for different timezones. Meanwhile, the IETF follows a model of aligning to the timezone of the hosting venue. It is unclear which model is better, since in some cases the focus is more on forming consensus and less on presentations. Though, collaboration that comes more naturally with in-person events becomes more difficult in online-only mode. One option is for online conferences to span several weeks with shorter, e.g., two hours per day, slots. The focus can also be shifted towards an online interim (topical) meeting rather than one or two broader events per year.

\section{Lessons Learned}
\label{subsec:lessons-online-everything}
We now present the summary of the views of groups 1 and 2, which discussed their experiences organising and participating in online conferences, meetings, teaching, etc. We split their assessment into two categories: What worked well and what did not. In both categories, we highlight the different topics, i.e., types of meetings, they commented on. We summarise the outcome of the group discussion in Tables I and II, reflecting the groups' conclusions. %about aspects of different online meetings and the necessary tooling to make them successful. %~\ref{tab:characteristics}

\subsection{Group 1}

\begin{table}[h!]
\label{tab:tools}
\begin{center}
\begin{tabular}{|l|c|c|c|c|c| }
 \hline
 \textbf{Type of Meeting} & \rot{\textbf{Audio}} & \rot{\textbf{Video}} & \rot{\textbf{Text channel}} & \rot{\textbf{\parbox{2.25cm}{Social interaction, \\ unstructured}}} & \rot{\textbf{\parbox{2.25cm}{Interactive, \\ e.g., dialog}}} \\
 \hline
PhD defense & \tableY & \tableY & \tableN & \tableN & \tableN \\
Conference & \tableY & \tableY & \tableY & \tableY & \tableN \\
Standardisation & \tableY & \tableP & \tableP & \tableN & \tableY \\
Panel/plenary discussion & \tableY & \tableY & \tableY & \tableN & \tableN \\
TPC & \tableY & \tableY & \tableP & \tableP & \tableY \\
Teaching & \tableY & \tableP & \tableY & \tableN & \tableN \\
Project meeting & \tableY & \tableY & \tableN & \tableN & \tableY \\
Tutorial & \tableY & \tableP & \tableY & \tableN & \tableN \\
Posters and Demos & \tableY & \tableP & \tableN & \tableY & \tableY \\
Research visits & \tableY & \tableY & \tableN & \tableY & \tableY \\
 \hline
\end{tabular}
\end{center}
\caption{Recommended supporting tools for various meeting types.~\tableY:~Strongly required,~\tableP:~Partial, and~\tableN:~Irrelevant}
\vspace{-12mm}
\end{table}

\textbf{What did not work well online} -- 
Online {teaching} sets higher requirements to students when it comes to self-management and dedication, which means that the dropout rate can be rather high, i.e., the number of participants decline over time. In a live lecture delivered as video stream (not pre-recorded), it is usually hard to capture when participants get lost, where some approaches such as text channels for Q\&A, e.g., Slack, exist though. As such, having a dedicated channel for posting questions, e.g., Slack or \textit{Tweetback}, that can be later addressed by the lecturer was perceived to work well, see Table I. This, however, requires further resources to handle the questions, since it is challenging for a lecturer to deliver the content and simultaneously follow the chat. Looking at {Technical Program Committee (TPC)} meetings, if there is no travelling associated, researchers tend to overcommit, where TPC meetings are usually difficult to squeeze into the overall schedule. This is simply a matter of habit and not an issue with online meetings per se. At an in-person TPC meeting, participants with conflict need to temporarily leave the room, i.e., every few minutes TPC members leave and re-enter the room, see Table II. 

\textbf{What worked well online} -- {TPC} meetings for lower-tier venues were held online or via the phone. Top-tier venues had, traditionally, in-person meetings that are now also held online. They work very well when everyone is prepared for the meeting, see Table II. With online TPC meetings, handling conflicts of interest is easier, where conflicted participants are sent to a breakout rooms and easily moved back, see Table I. Meanwhile, accessibility of online TPC meetings has (and should be) increased since no travelling is required and no dedicated budget. 

Also, {research visits} are another variation of meeting, which led to in-person group meetings and day-to-day discussions, see Table II. There is now also vast experience in online {teaching} available: We noted that online teaching can generally lead to multiple outcomes. First, better grades, since participants that take the exam are usually motivated. If recordings are provided, students have always access to an explanation of the content. Yet, many participants believe that asynchronous teaching material, e.g., videos, will be the future, e.g., explanations of an algorithm can be viewed multiple times, as previously mentioned. The most difficult part are definitely lab sessions with need to hardware access. In all other cases, virtualization and remote access work well, see Table I. 

Finally,, we considered two categories of {project meetings}, see Table II: \emph{Administrative} such as general assembly or EU project review meetings, commonly in Brussels. Having these meetings online has not made things worse. The second kind of meetings are \emph{preparatory} with intention to get the project started such as to get teams set up, build community (no social activities, but still due to different contacts). To this end, what helps is the social need that participants have to get to know each other and move things forward, although such interactions are very participant dependent.

\begin{table*}[h!]
\begin{center}
\label{tab:characteristics}
\begin{tabular}{ |l|c|c|c|c|c| }
 \hline
 \textbf{Type of Meeting} & \rot{\textbf{\parbox{2.4cm}{Preparation/ \\ Efforts}}} & \rot{\textbf{Agenda-driven}} & \rot{\textbf{\parbox{2.4cm}{Goal/ \\Problem-driven}}} & \rot{\textbf{\parbox{2.4cm}{Creativity, \\ Unstructured \\ actions}}} & \rot{\textbf{\parbox{2.4cm}{Unstructured \\ actions beyond \\ agenda}}} \\
 \hline
PhD defense	& Same & \tableN & \tableY & \tableN & \tableY{} PhD celebration \\
Conference & Demanding & \tableY & \tableN & \tableN & \tableY{} social event, hallway discussions \\
Standardisation & Demand & \tableY & \tableY & \tableN & \tableY{} social event, hallway discussions \\
Panel/plenary discussion & Same & \tableN & \tableN & \tableY & \tableY{} social event, small group discussions \\
TPC & Same & \tableY & \tableY & \tableN & \tableP{} \\
Teaching & Intense & \tableY & \tableY & \tableY & \tableN{} \\
Project meeting	& Same & \tableY & \tableY & \tableN & \tableP{} social event, small group of interest \\
Tutorial & Same & \tableY & \tableY & \tableN & \tableP{} small group of interest \\
Posters and Demos & Demanding & \tableY & \tableY & \tableN & \tableY{} social event, hallway discussions \\
Research visits & Same & \tableN & \tableN & \tableY & \tableY{} daily activities \\
 \hline
\end{tabular}
\end{center}
\caption{Meeting requirements for various meeting types. ~\tableY:~A particular characteristic of the meeting is a strong requirement for its success,~\tableP:~Partial, and~\tableN:~Irrelevant}
\vspace{-10mm}
\end{table*}

\subsection{Group 2}
\label{subsec:lessons-online-everything}

\textbf{What did not work well online} -- Online {PhD defenses} reach their goal, but lack the celebration aspects, see Table II. When it comes to {teaching}, pre-recording lectures do require an enormous effort. Teaching also feels as if it is performed into the void, without perceiving reactions, e.g., if the presented content is understood or the students are bored or lost, see Table II.
    
{Online meetings} on the other hand face their own issues: They can generate churn as participants join and leave. Participants might be absent, leaving their computer connected, making it hard to identify who is actually present. There is also a tendency for people to over-do/commit in the number of online meetings. Too many meetings lead to fragmentation and eventual loss of context. In addition, without proper calendar invites, finding meeting information, e.g., links, passwords, etc., can also become cumbersome. Timezones further complicate scheduling and limit available meeting options. Generally, it is difficult to quantify ``missed opportunities''. It seems that attempts to simulate the in-person experience does not work.

Current {online tools} provide no way to capture social cues. For instance, when it would be acceptable to interact with participants and when not. This is very easy in an in-person setting and currently impossible online, see Table I. During in-person meetings, one typically talks to their neighbours, while in online meetings, everyone is a neighbour. As such, the question is whom should you talk to? There is currently an excess in the number of online tools available: Where do we meet?, What is a shared platform that everyone has installed?. When scheduling meetings, this information needs to be captured along the available time slot(s), making meeting scheduling more cumbersome. It remains unclear whether this increases the cognitive load.
    
When it comes to {social activities}, "forced fun-on-demand" is challenging, see Table I. Social activities work if they are prepared, e.g., a birthday party where wine is shipped to everyone or ingredients for the social event are made available beforehand. This is, however, very participant dependent.

%\item
%  Exhibition / presentation booths to set context for discussions can
%  also work online

\textbf{What worked well online} -- Online {Technical Program Committee (TPC)} meetings seem to work well and can be handled very efficiently, see Table II. Group work is performed well by using random assignments to breakout rooms. In this mode, participants are assigned to smaller groups, e.g., up to four people, at random by using breakout rooms. There are two examples. First, (panel) discussions in smaller groups, e.g., IEEE QoMEX 2021, and Dagstuhl style group discussions. Secondly, getting to know new people by randomly assigning conference participants to smaller breakout rooms works well. Project meetings worked well, since people know each other. Maybe more productive online than in-person since discussions are more focused with fewer disruptions. The downside here however is that people tend to overcommit. {Interactive discussion} with speakers during talks also work well, but may lead to exhaustion, if they run for too long, see Table I. Stopping by a conference for just a single session is possible in online conferences, since no travel is needed, which is a real benefit of virtual conferences. Meanwhile, {pre-recorded presentations} become part of the proceedings and are although (mostly not as permanently) archived just like research papers, which is a real benefit for the scientific community. {Q\&A sessions} work better in online mode, where more questions are being asked by junior participants. {Shared edition of reports} is also possible; in an in-person meeting, it is typically considered a bad habit to use a laptop during the meeting, so online note taking is less common. Online mode also opens {new meeting opportunities} since it is very cheap (also time-wise) to interact with new communities that one normally would not attend. Online birthday parties can also work - for example by ordering a bottle of wine or pizza to each participant - same wine and food for everyone creates a joint experience. Playing online interactive games, e.g., escape room, can also provide an immersive real-world experience.

\section{Takeaways} 

The COVID-19 pandemic brought more attention to the need of making typically in-person activities in the research community more accessible. After the assessment of all participants of the Dagstuhl seminar on ’Climate Friendly Internet Research’, we arrived at some conclusions in Section \ref{subsec:lessons-online-everything} that we would like to present in the following.

Firstly, different types of meetings have different requirements and audiences. If online meetings are successful, it mainly depends on their requirements. For instance, certain meetings achieve their goals if the agenda is fulfilled, and thus can work very well online. Other meetings have important goals beyond a specific agenda, e.g., important activities that are triggered in hallway conversations and random encounters and not only by \textit{ticking off} items from a list or content dissemination.
As such, it is also important to be goal-oriented, e.g., structured activities have shown to work well online when the online tools meet the needs of the meeting. For example, when it comes to the IETF, there is a non-negligible investment in the tailored tooling for the needs of the community, which offered hybrid participation even before the pandemic. Meanwhile, unstructured activities, e.g., whiteboard sessions, random encounters and discussions during coffee breaks or hallway conversations, have shown to do not work well. While online meetings are more focused, having fewer distracting factors, they lack the overall social cues from the audience. With these aspects in mind, online conferences are not meant to fully replace every aspect of in-person conferences. They have however to provide the same framework for the participants, as mentioned in Section~\ref{sec:guidelines:and:best:practices}.

Secondly, all participants agreed that online meetings \emph{can} challenge work-life balance. Preparing online teaching material in general takes more time, where some participants reported up to ten times as long compared to in-classroom teaching. In general, all participants reported that their work became more intensive, since more meetings are being scheduled for things that could have been solved in-person with a quick, informal, discussion. This so-called meeting overload is reflected in the typical gap between meetings: The gap between in-person meetings is usually around five minutes (or more), online meetings, on the other hand, is often less as many meetings start and end on the hour. %Consequently, this more intense schedule can lead to health stressing factors since people move less, e.g., do not leave their chair for ten hours.

Finally, we conclude that online conferences will never fully replace every aspect of an in-person conference, in particular when it comes to face-to-face (small group and individual) so-called hallway discussions. Nevertheless,  we identified that they may actually be better than in-person conferences found today: Widening participation, where in-person conferences have been classically strict about, e.g., when issues such as visa, health, family, budget or air travel restrictions arise and none of the authors can be present, see Section~\ref{subsec:implications-online-conferences}.

% \vspace{-15mm}
% \begin{IEEEbiographynophoto}
% 	{Simone Ferlin}
% 	is a senior performance engineer at Red Hat. She received her PhD degree in computer science from the University of Oslo, Norway in 2017. Her interests lie in the intersection of cellular networks and the Internet, focusing on QoS and cross-layer design, transport protocols, congestion control, and network measurements.
% \vspace{-15mm}
% \begin{IEEEbiographynophoto}
% 	{Oliver Hohlfeld}
% 	is a Professor and Chair of Computer Networks at Brandenburg University of Technology (BTU). Before, he headed the Network Architectures group at RWTH Aachen University. He obtained a Ph.D. from TU Berlin in 2013.
% \begin{IEEEbiographynophoto}
% 	{Vaibhav Bajpai}
% 	is an independent research group leader at CISPA Helmholtz Center for Information Security. He received his PhD (2016) and Masters (2012) degrees from Jacobs University Bremen.
% \vspace{-15mm}

%a general question we asked is how to lower friction? For instance, lower friction activities happen easily online, while higher friction activities are missed. Friction can also be increased artificially, where one can consider fetching and sending emails only once per day, thus, increasing the turnaround for a response to an email noticeable to others. %and thereby helps to focus on getting work done. 

% \subsubsection{Synthesis: What did we learn? \& General recommendations}

\bibliographystyle{IEEEtran}
\bibliography{Bibliography}

\end{document}